\documentclass[runningheads]{llncs}
\usepackage[T1]{fontenc}
\usepackage{graphicx}
\usepackage{amsmath,amssymb,amsfonts}
\usepackage{booktabs}
\usepackage{array}
\usepackage{url}
\usepackage{tikz}
\usetikzlibrary{arrows.meta,positioning}
\usepackage[hidelinks]{hyperref}

\begin{document}

\title{CCR: Towards a Common, Quality-Gated CACAO Integrations Registry for European Cybersecurity Automation\thanks{This work is partially funded by the European Union. Views and opinions expressed are however those of the author(s) only and do not necessarily reflect those of the European Union or Smart Networks and Services Joint Undertaking. Neither the European Union nor the granting authority can be held responsible for them. The XTRUST-6G project (GA no. 101192749) is supported by the 6GSNS and its members.}}

\titlerunning{CCR: An Open, Quality-Gated Registry of CACAO Integrations}

\author{Mateusz Zych\inst{1,2} \and Vasileios Mavroeidis\inst{2} \and Gudmund Grov\inst{3,2}}

\authorrunning{M. Zych et al.}

\institute{Cyentific AS, Oslo, Norway\\
\email{zych@cyentific.eu} \and
Department of Informatics, University of Oslo, Oslo, Norway\\
\email{\{mateusdz, vasileim, gudmungr\}@ifi.uio.no} \and
Norwegian Defence Research Establishment, Kjeller, Norway\\
\email{Gudmund.Grov@ffi.no}}

\maketitle
\begin{abstract}
Standardised, machine-readable cybersecurity playbooks provide a basis for portable, shareable, and reusable incident-response logic. OASIS CACAO provides a vendor-neutral representation for such playbooks, but not the product-specific integration artefacts needed to invoke external products and services. We introduce the Common CACAO Registry (CCR), an open, provenance-aware registry of CACAO HTTP-API connector envelopes. Each envelope captures an API operation's command, inputs, target, authentication-related information, provenance, validation evidence, and maturity metadata. CCR is \emph{quality-gated}, with acceptance requiring both CACAO v2 schema validity and a mean back-validation score of at least 0.8 against the source OpenAPI operation, while a six-level maturity model records progressively stronger evidence and distinguishes gate acceptance from operational readiness. To seed CCR, we develop a hybrid OpenAPI-to-CACAO pipeline. Deterministic code extracts source-derived interface facts, generates identifiers, wires cross-references, and validates structure, while a constrained LLM provides bounded semantic enrichment, including action naming, authentication interpretation, and CACAO activity annotation. Evaluation across eight security APIs yields 713 CACAO-schema-valid envelopes with a mean back-validation score of 91.5\%, of which 675 produce well-formed, dispatchable HTTP requests in a local harness. Comparison with a deterministic rule-based baseline shows that mechanical API structure is preserved more reliably through rule-based translation, while the LLM contributes bounded semantic enrichment, most notably CACAO activity annotation. Together, these results support CCR as reusable integration infrastructure for CACAO action steps and as an initial foundation for a broader common European registry.

\end{abstract}

\section{Introduction}
Security operations teams coordinate detection, investigation, containment, remediation, recovery, communication, compliance, and reporting through documented incident-response procedures \cite{islam2019multivocal}. A \emph{response playbook} applies such a procedure to a defined incident scenario and captures the sequence and conditional branching of response activities. Some activities are performed manually, others are supported or automated by security tools, and recurring sequences of activities can be orchestrated through Security Orchestration, Automation, and Response (SOAR) and related response platforms.

In a machine-readable response playbook, response activities can be represented as action steps. When an action step is automated through an external tool, it must be connected to a concrete capability exposed by a product, service, or data source deployed in an organisation's environment. An action such as retrieving a vulnerability record, creating a case, blocking an indicator, or changing a network-policy rule therefore requires a usable technical binding comprising the target interface, communication protocol, operation or command, variables, authentication information, and other details needed to invoke the relevant capability. These bindings must be created, reviewed, maintained, and adapted as security products and their interfaces evolve. Therefore, in heterogeneous or large-scale environments, the effort required to build and maintain these integrations can become a major bottleneck for operational automation.

OASIS CACAO provides a vendor-neutral, machine-readable representation for cybersecurity playbooks, including workflow constructs, action steps, targets, variables, authentication-related objects, and, among other command types, HTTP-API commands \cite{cacao2023}. A CACAO playbook can represent response logic in a structured form, including the sequence, branching, and execution-relevant details of manual or automated activities \cite{mavroeidis2021integration,gurabi2022sasp,tsirakis2025operationalizing}. CACAO therefore supports the documentation, sharing, and reuse of selected response logic across organisational and technological boundaries. However, playbook authors and automation environments still need access to reusable, inspectable product-specific bindings that can connect action steps to the concrete capabilities of the tools they invoke.

Many security products expose such capabilities through HTTP APIs described using OpenAPI, which specifies operations, parameters, request bodies, servers, and security schemes in a machine-readable form \cite{openapi2021,casas2021openapi}. This work focuses on this class of integrations. OpenAPI specifications provide structured evidence from which CACAO-side integration artefacts can be derived, but the mapping involves both mechanical and semantic decisions. Interface facts, such as HTTP methods and declared parameters, should be preserved faithfully, whereas other decisions concern how an operation should be named and exposed to a playbook, how its inputs should be represented, and what operational purpose the action serves. In this work, semantic enrichment includes annotation with CACAO's \texttt{playbook\_activity} vocabulary, which gives eligible actions a common operational label and supports future discovery, comparison, and interpretation across tools.

This paper introduces the \emph{Common CACAO Registry} (CCR), an open, versioned, provenance-aware registry of CACAO HTTP-API connector envelopes. Each envelope captures the reusable, action-level representation of one API operation, along with source-specification provenance, validation evidence, assurance metadata, and maturity information. A response playbook can combine multiple such action steps into a workflow, while adopting organisations complete the local configurations needed for their products, credentials, endpoints, policies, and approval processes. CCR provides an initial common foundation through which contributors can generate, submit, review, correct, and maintain integration artefacts under a shared quality framework. Its open licensing, standards-based representation, and public contribution model make integration knowledge inspectable, reusable, and jointly maintainable across a wider cybersecurity ecosystem.

To seed CCR, we develop a five-stage OpenAPI-to-CACAO generation pipeline that assigns structurally deterministic tasks---including identifier generation, cross-reference construction, and schema validation---to code, while using a constrained large language model (LLM) for bounded semantic decisions, including action naming, variable enrichment, authentication interpretation, and activity annotation. We evaluate this pipeline against a deterministic rule-based baseline to identify an appropriate division of labour between mechanical translation and semantic enrichment. This evaluation is guided by the following research questions:

\begin{description}
    \item[RQ1 (division of labour):] Which decisions in OpenAPI-to-CACAO translation benefit from a constrained LLM, and which does a deterministic, rule-based mapper handle at least as well? \label{sec:rq1}
    \item[RQ2 (execution-gate validity):] To what extent can the registry's automated quality gate be trusted under execution-oriented checks independent of the OpenAPI specification from which the artefacts were derived? \label{sec:rq2}
\end{description}

CCR also sits within a broader European regulatory and policy context concerning incident handling, resilience, vulnerability management, coordinated response, interoperability, and reusable digital capabilities \cite{nis2directive,cerdirective,craregulation,cybersolidarityact,interoperableeuropeact}. Section~\ref{sec:background} develops these connections and their relevance to shared cybersecurity integration infrastructure. Against this background, the contributions of this paper are fourfold:

\begin{itemize}
    \item We specify and release CCR: an open, standards-conformant registry of CACAO integration artefacts with explicit quality gates, a six-level artefact maturity model, and an open contribution model.
    \item We seed it with 713 CACAO-schema-valid connector envelopes across eight real-world security APIs, produced by a prototype generation pipeline.
    \item We provide a layered validation framework that combines schema validation, back-validation, a deterministic baseline, dispatch testing, and limited pipeline execution, stating explicitly what each layer does establish.
    \item We connect this infrastructure to European cybersecurity regulation and policy, arguing that a shared integration registry can reduce duplicated engineering effort for capabilities relevant to NIS2 and wider EU resilience and interoperability objectives.
\end{itemize}

The remainder of the paper motivates the integration problem and its European context, presents the OpenAPI-to-CACAO generation pipeline and CCR design, evaluates the resulting integration corpus against the research questions, positions the contribution relative to related work, discusses its implications and limitations, and concludes with next steps towards a broader operational registry.

\section{Motivation and Background}\label{sec:background}
SOAR platforms coordinate and automate actions across heterogeneous security systems such as SIEMs, EDR platforms, firewalls, ticketing systems, and threat-intelligence services \cite{islam2019multivocal}. Their practical utility depends not only on the expressiveness of response workflows but also on the availability, correctness, and maintainability of the integrations that enable those workflows to invoke external tools. Security-tool integration has been identified as a time-consuming and resource-intensive part of SOAR engineering \cite{islam2020architecture}. The problem persists after an integration is created because connectors encode assumptions about interfaces, parameters, authentication, and other API behaviour that may change over time. In a study of 2,224 OpenAPI specifications, 251 API versions introduced breaking changes, and 87.3\% of these did not first deprecate the superseded operation \cite{9240687}. Consequently, manually creating and maintaining product-specific integrations becomes increasingly difficult as the number and diversity of security tools grow.

Within the HTTP-API scope of this work, integration can be viewed as a translation between representations with different purposes. OpenAPI describes service operations and interface elements primarily from the provider's perspective \cite{openapi2021,casas2021openapi}, whereas CACAO describes actions from the playbook and orchestration perspective \cite{cacao2023}. HTTP methods, paths, and declared parameters can often be derived mechanically, whereas representing variables, authentication information, targets, and operational semantics requires additional decisions. A reusable integration therefore requires a CACAO-conformant binding that preserves source-derived interface facts while expressing the operation in a form suitable for incorporation into a cybersecurity playbook.

CCR is also motivated by the wider European regulatory and policy context. NIS2 establishes cybersecurity risk-management measures covering incident handling, supply-chain security, and vulnerability handling and disclosure, together with time-bound reporting obligations for significant incidents \cite{nis2directive}. The CER Directive, Cyber Resilience Act, Cyber Solidarity Act, and Interoperable Europe Act broaden this context through requirements and policy objectives concerning resilience, product cybersecurity, coordinated response, interoperability, and reuse of digital capabilities \cite{cerdirective,craregulation,cybersolidarityact,interoperableeuropeact}. The ENISA Threat Landscape 2025 further reports vulnerability exploitation as the second-leading initial-intrusion route and increased targeting of critical dependencies in digital supply chains \cite{enisa2025threatlandscape}. These findings reinforce the relevance of reusable, transparent, and interoperable infrastructure for coordinating incident response across heterogeneous security products, without prescribing any particular platform, standard, or registry.

ENISA's European Vulnerability Database (EUVD) \cite{enisaeuvd} provides a useful, although functionally distinct, precedent for shared cybersecurity infrastructure by aggregating vulnerability information that would otherwise be reconstructed independently. CCR applies the same sharing principle at the integration layer, where reusable integration knowledge can be maintained collectively across organisations and automation environments. 

Table~\ref{tab:nis2map} maps selected EU regulatory and policy objectives to CCR contributions in areas with a direct technical relationship to an integration registry.

\begin{table}
\caption{Mapping selected EU regulatory and policy objectives to CCR capabilities.}
\label{tab:nis2map}
\centering
\scriptsize
\setlength{\tabcolsep}{3pt}
\begin{tabular}{>{\raggedright\arraybackslash}p{1.9cm}p{3.8cm}p{5.5cm}}
\toprule
\textbf{EU Instrument/initiative} & \textbf{Relevant objective} & \textbf{CCR contribution} \\
\midrule
NIS2 Art.~21(2)(b) \cite{nis2directive} & Entities must implement ``incident handling'' among minimum risk-management measures. & The executable, vendor-specific binding layer (validated CACAO action steps) that lowers the integration burden of automating a documented incident-handling procedure. \\
NIS2 Art.~23(1),(4) \cite{nis2directive} & Early warning within 24h, notification within 72h of a significant incident. & Lowers the fixed cost/time of building automatable detection-to-triage workflows relevant to meeting these deadlines. \\
NIS2 Art.~21(2)(e) \cite{nis2directive} & Measures must cover vulnerability handling and disclosure. & Seed content includes back-validated candidates for vulnerability-intelligence sources (e.g., CVE Services API). \\
NIS2 Art.~21(2)(d); CRA \cite{nis2directive,craregulation} & Supplier/service-provider risk management; product vulnerability-handling duties. & Openly licensed, auditable registry with a transparent contribution model reduces single-vendor dependency in the SOAR tooling supply chain. \\
CER; Cyber Solidarity Act \cite{cerdirective,cybersolidarityact} & Physical/operational resilience duties; EU cross-border incident coordination (Cyber Hubs). & Same ``shared rather than duplicated'' logic applied at the integration layer, as EUVD \cite{enisaeuvd} already applies to vulnerability data. \\
Interoperable Europe Act; Horizon Clusters 3/4 \cite{interoperableeuropeact,horizonccl3eccc2026,horizoncl4openinternetstack2026} & Public-sector reuse of interoperability solutions (incl.\ source code); EU R\&I direction towards open, standards-based security automation. & Open licensing, standards-based artefacts, and a public contribution model mirror the pattern these instruments require or fund. \\
\bottomrule
\end{tabular}
\end{table}

\section{The CCR Generation Pipeline}\label{sec:design}

Translating OpenAPI operations into CACAO integrations admits three design approaches. A \emph{rule-based} mapper is deterministic and computationally inexpensive and is well suited to preserving structurally explicit API information. However, deriving higher-level operational semantics from endpoint names, descriptions, parameters, and surrounding context is substantially harder to capture through fixed rules alone and would require increasingly complex handcrafted heuristics or additional language-processing components. A \emph{pure-LLM} mapper could generate a complete connector envelope in a single step, but would also place identifiers, cross-references, and structural correctness under model control, despite these aspects being more reliably handled deterministically.

\begin{figure}[t]
\centering
\includegraphics[width=\textwidth]{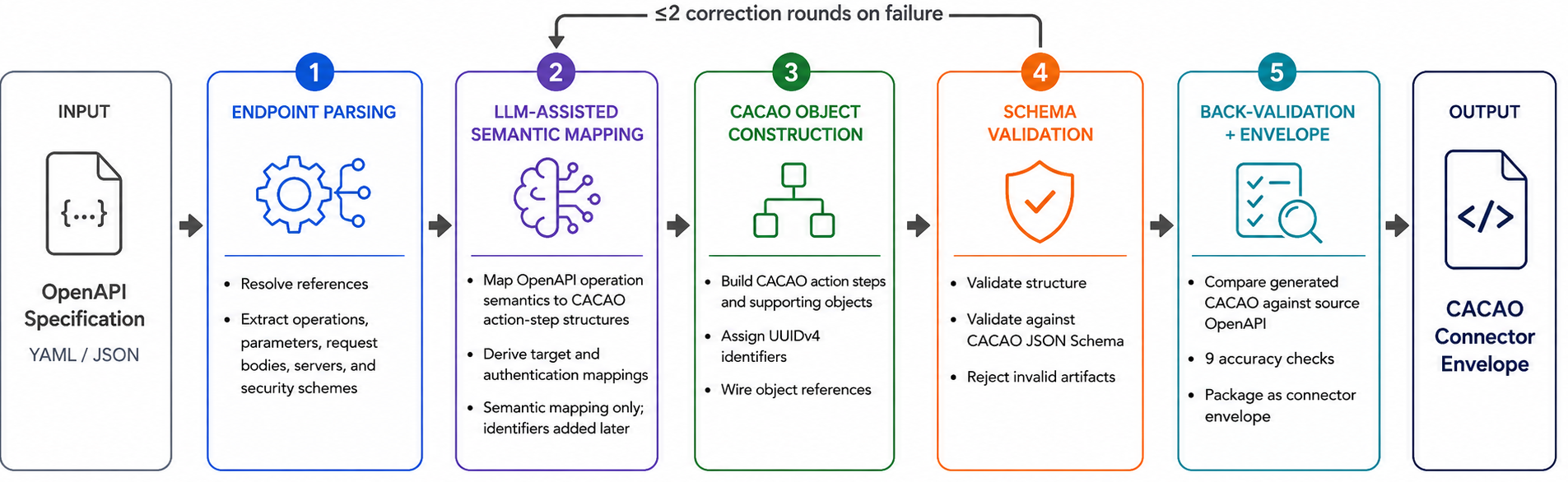}
\caption{Five-stage OpenAPI-to-CACAO pipeline used to generate the connector envelopes seeded into CCR.}
\label{fig:pipeline}
\end{figure}

Our pipeline follows a hybrid design, combining an AI-assisted five-stage process with deterministic identifier generation, cross-reference construction, and validation, while reserving the LLM for bounded semantic decisions not directly determined by the OpenAPI specification, including variable naming, type enrichment, and vocabulary-grounded activity annotation. We refer to this prototype OpenAPI-to-CACAO generator as \emph{CACAO Bridge}, or simply Bridge in the evaluation. Table~\ref{tab:det_vs_llm} summarises this division of labour, while RQ1 evaluates the contribution of the LLM relative to the deterministic rule-based alternative. Figure~\ref{fig:pipeline} illustrates the pipeline. Although shown linearly, Stage~4 can return to Stage~2 for up to two correction rounds.

\begin{table}[t]
\caption{Deterministic vs.\ LLM-handled decisions in the pipeline.}
\label{tab:det_vs_llm}
\centering
\scriptsize
\setlength{\tabcolsep}{3pt}
\begin{tabular}{p{5.5cm}p{1.7cm}p{4cm}}
\toprule
\textbf{Decision} & \textbf{Handler} & \textbf{Why} \\
\midrule
HTTP method/path template extraction & Code & Single field in OpenAPI \\
$\$ref$/allOf resolution & Code & JSON pointer lookup \\
UUID and object ID generation & Code & Requires uniqueness guarantee \\
Cross-reference wiring & Code & Structural, no ambiguity \\
JSON schema \& back-validation & Code & Deterministic comparison \\
\midrule
Action step name/description & LLM & Synthesized from multiple fields \\
Command string, variable naming/typing & LLM & Naming and semantic judgment \\
Required/external flag, header placement & LLM & Cross-standard alignment \\
Authentication type mapping & LLM & Cross-vocabulary alignment \\
Playbook activity annotation & LLM & Controlled-vocabulary judgment \\
\bottomrule
\end{tabular}
\end{table}

\textit{Stage 1 - Endpoint extraction:} The input OpenAPI document is parsed and all operations are extracted. OpenAPI specifications commonly factor out shared definitions using \texttt{\$ref} pointers and describe composite or alternative request bodies with the \texttt{allOf}/\texttt{anyOf}/\texttt{oneOf} schema-composition keywords. Stage 1 resolves both into a single, self-contained, normalised endpoint record, so later stages never have to re-derive an indirect reference.

\textit{Stage 2 - LLM mapping:} The normalised endpoint record is sent to a Claude model, using \texttt{claude-haiku-4-5} for simple endpoints and \texttt{claude-sonnet-4-6} for endpoints with at least five parameters or a request body; this routing affects inference cost but not the pipeline structure. The model receives CACAO v2 context together with CACAO's \texttt{playbook-activity-type-ov}, an open controlled vocabulary of operational-purpose labels such as \texttt{scan-system} and \texttt{block-ip} that allows an action step to express its operational purpose independently of the underlying HTTP command \cite{cacao2023}. Structured tool use constrains the model to a typed output schema covering names, command strings, variables, targets, authentication semantics, and \texttt{playbook\_activity}. The model therefore produces only the semantic content required for subsequent construction, while identifiers remain under deterministic control.

\textit{Stage 3 - Deterministic object construction:} A \texttt{Node.js} builder assigns CACAO-conformant \texttt{type--UUIDv4} identifiers, wires cross-references between the action step, target, and authentication-info objects, and assembles the connector envelope, keeping identifier generation entirely outside the LLM's control to prevent duplication or format errors.

\textit{Stage 4 - Schema validation and self-correction:} The envelope is validated against the CACAO v2 JSON Schema family using Ajv, a standard JSON Schema validator. On failure, the structured error list is fed back to the model for up to two correction rounds before an endpoint is flagged for manual review.

\textit{Stage 5 - Semantic back-validation:} Passing Stage 4 only proves the envelope is well-formed and schema-conformant; a schema-valid envelope with the wrong HTTP method, a swapped parameter type, or a missing authentication object would pass Stage 4 undetected. Stage 5 closes that gap by comparing the generated envelope directly against the source OpenAPI operation on nine dimensions, defined and reported in Section~\ref{sec:evaluation}.

The unit of generation is a single \emph{connector envelope} per endpoint, represented as a self-contained JSON document containing the CACAO objects for one API operation together with back-validation scores and generation telemetry. CCR stores connector envelopes, while complete playbooks can compose multiple action steps into executable workflows. The public artefact package includes the generated envelopes, OpenAPI specifications and hashes, validation evidence, and execution demo, while the prototype generator itself is not released as a reusable contribution.\footnote{CCR Repository: \url{https://github.com/cacao-common-registry/integrations}. Later artefact paths are relative to it.}

\subsection{Worked Example}\label{sec:workedexample}
The worked example uses the CVE Services operation for retrieving a CVE record by identifier, which is subsequently used in the live-execution evaluation. The resulting connector envelope is available in the registry.\footnote{\texttt{integrations/cve-services-api/2.5.2/get\_cve-id.json}} Stage 1 extracts the HTTP method, path, and single required string identifier parameter. The operation defines no request body, security requirement, or schema composition requiring resolution. Stage 2 names the action, writes the command with the identifier as an externally supplied CACAO variable, assigns an activity label, and correctly emits no authentication object. Stage 3 assigns UUIDs outside the LLM. Stage 4 validates the envelope on the first attempt, and Stage 5 passes all nine applicable back-validation checks.

\textit{Schema composition handling.} Request bodies that combine sub-schemas via the \texttt{allOf}/\texttt{anyOf}/\texttt{oneOf} keywords introduced above (Stage 1) are flattened into one combined set of fields, a bounded procedure limited to depth 6. This treatment is exact for \texttt{allOf}, but constitutes an over-approximation for \texttt{anyOf}/\texttt{oneOf}. Under \texttt{anyOf}, an instance must match at least one alternative branch, whereas under \texttt{oneOf}, it must match exactly one. The pipeline nevertheless unions the fields from all branches rather than distinguishing between the alternatives, since it does not consult OpenAPI \emph{discriminator} fields. The same treatment applies when an operation declares multiple request-body content types. OpenAPI callbacks are not supported, and operations that depend on them remain outside the pipeline's scope.

\section{The Common CACAO Registry}\label{sec:registry}
CCR is a public, versioned repository of CACAO HTTP-API connector envelopes, organised by vendor, product and version and structured for CACAO-conformant orchestration engines. The evaluation in this work establishes CACAO conformance and direct dispatch of the underlying HTTP requests, while end-to-end integration with a third-party orchestration engine remains outside the present evaluation.

\textit{Structure.} One connector-envelope JSON file is provided per OpenAPI endpoint, grouped under \texttt{integrations/\textless{}product\textgreater{}/\textless{}version\textgreater{}/}, alongside source OpenAPI documents, SHA-256 hashes, validation evidence, an assurance schema, and a per-API manifest.

\textit{Quality gates.} A submission is accepted only if it (i) passes CACAO v2 schema validation and (ii) achieves an overall back-validation score of at least \emph{0.8}. The threshold was selected empirically with reference to the score distribution of the seed corpus; under the nine-check scoring scheme, it permits one failed check and a partially satisfied second check, provided the remaining seven pass. At this threshold, 82.5\% (588/713) of the envelopes qualify for acceptance, whereas the remainder are retained with a \emph{needs-review} status. The resulting gate is subsequently evaluated under RQ2 using execution-oriented evidence independent of the source-specification comparison.

\textit{Contribution model and licensing.} CCR supports two contribution paths. Contributors may submit an independently generated envelope or manually author or correct an envelope directly. Both paths converge on the same representation and are subject to the same acceptance criteria, allowing contributors to add and maintain integrations independently of the original authors. Registry artefacts, validation scripts, schemas, copies of the source specifications, and demonstration materials are released under the Apache-2.0 license, supporting the digital-sovereignty objective discussed earlier.

\textit{Artefact package.} The package contains all 713 envelopes, source OpenAPI files and hashes, per-API manifests, the assurance schema, validation outputs, and a minimal CACAO playbook demonstration assembled from CCR envelopes.

\textit{Seed content.} At the time of writing, CCR is seeded with the 713 connector envelopes evaluated in Section~\ref{sec:evaluation}, spanning eight security APIs across vulnerability intelligence, case management, SOAR, network policy, and SIEM/EDR domains (Table~\ref{tab:dataset}). The significance of the seed corpus lies not in its size relative to existing connector collections, but in the common representation and quality framework applied across its contents. Each of the 713 artefacts is CACAO-schema-valid, has recorded back-validation evidence, and is openly licensed under a common standard.

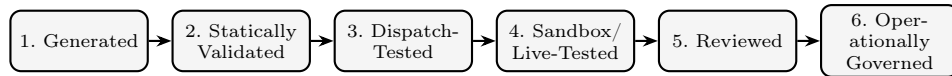
\begin{figure}
\centering
\begin{tikzpicture}[
  stage/.style={rectangle, draw, thick, rounded corners, text width=1.6cm, minimum height=.8cm, align=center, font=\scriptsize, fill=gray!8},
  arr/.style={-{Stealth[length=2mm]}, thick}
]
\node[stage] (n1) {1. Generated};
\node[stage, right=3mm of n1] (n2) {2. Statically\\Validated};
\node[stage, right=3mm of n2] (n3) {3. Dispatch-\\Tested};
\node[stage, right=3mm of n3] (n4) {4. Sandbox/\\Live-Tested};
\node[stage, right=3mm of n4] (n5) {5. Reviewed};
\node[stage, right=3mm of n5] (n6) {6. Operationally\\Governed};
\draw[arr] (n1) -- (n2);
\draw[arr] (n2) -- (n3);
\draw[arr] (n3) -- (n4);
\draw[arr] (n4) -- (n5);
\draw[arr] (n5) -- (n6);
\end{tikzpicture}
\caption{CCR's six-level artefact maturity model. Each envelope is tagged with the highest level its actual evidence supports.}
\label{fig:maturity}
\end{figure}

\textit{CCR as a governance-aware automation supply chain.}\label{sec:governance}
A connector-artefact registry is more than a passive document store. Once bound to operational endpoints and credentials, an envelope can trigger consequential actions, including destructive ones. We therefore treat CCR as part of an automation supply chain whose provenance, validation, contribution, and operational use require explicit governance. The current implementation focuses primarily on artefact-level governance through common contribution paths, acceptance criteria, CACAO schema validation, back-validation, and recorded maturity evidence. The maturity model in Figure~\ref{fig:maturity} provides a broader governance specification that separates these controls from progressively stronger levels of assurance based on dispatch testing, sandbox or live execution, review, and ultimately operational governance.

\begin{table}%[t]
\caption{Maturity levels and this paper's current evidence against each.}
\label{tab:maturitystatus}
\centering
\scriptsize
\setlength{\tabcolsep}{3pt}
\begin{tabular}{>{\raggedright\arraybackslash}p{3.3cm}p{8.5cm}}
\toprule
\textbf{Level (Figure~\ref{fig:maturity})} & \textbf{Current evidence (this paper)} \\
\midrule
1. Generated & 713/713 endpoints (Section~\ref{sec:evaluation}). \\
2. Statically validated & 713/713 schema-valid; 91.5\% mean back-validation score (Section~\ref{sec:results}). \\
3. Dispatch-tested & 675/713 (94.7\%); the other 38 are flagged with a specific defect rather than silently passed (Section~\ref{sec:execval}). \\
4. Limited live/mock-tested & CVE Services \texttt{GET /cve/\{id\}} has 5/5 live read-only requests; the playbook demo adds a local mock state-changing trace; seven APIs are not live-tested (Section~\ref{sec:execval}). \\
5. Reviewed & Not assessed in this study; consequently, no envelope currently claims this level. \\
6. Operationally governed & Not yet implemented; future work (Section~\ref{sec:governance}). \\
\bottomrule
\end{tabular}
\end{table}

CCR uses a six-level maturity model that classifies each artefact according to the strongest level supported by its available evidence (Figure~\ref{fig:maturity}). Each successive level requires stronger evidence, separating structural validity and back-validation from dispatchability, execution testing, review, and ultimately operational governance. In the current corpus, most envelopes reach Level~3 through successful dispatch testing, while dispatch failures remain at Level~2. The live-tested CVE operations and the playbook demonstration reach Level~4 only for the evidence established by those tests. Higher maturity levels are not implied by the corpus-wide mean back-validation score of 91.5\% and require their own supporting evidence.

\paragraph{Assurance envelope and threat model.} Each seed envelope carries an \texttt{assurance} block recording source provenance and its SHA-256 hash, maturity level and evidence reference, validation summary, credential hints, target URL constraints, action-impact class, review status, lifecycle state, and known limitations. These metadata support artefact-level assurance but do not constitute operational governance. Because CCR contains executable integration artefacts that may be contributed by multiple parties and ultimately invoke external APIs, its threat model includes malicious or erroneous target URLs, excessive credential scope, and incorrect classification of potentially destructive operations. The maturity model therefore reserves Level 6 for artefacts whose deployment and use are subject to operational governance controls. Establishing and evaluating those controls remains future work.

\paragraph{Example integration execution.}
CCR includes a two-action CACAO playbook demonstration.\footnote{\texttt{execution-validation/playbook-demo/}} The first action performs a live CVE Services \texttt{GET /cve/\{id\}} lookup. The second uses a TheHive alert-creation envelope to a local recording server, producing a mock state-changing trace. The assembled playbook passes CACAO JSON Schema validation, and execution through the local harness succeeds with HTTP 200 and HTTP 201 responses for the two actions, respectively. Although the playbook is not executed through a CACAO-enabled SOAR platform, the experiment demonstrates how CCR envelopes are composed into a multi-action playbook and the resulting executable HTTP request flow.

\section{Evaluation}\label{sec:evaluation}
We evaluate CCR's seed-content quality, and thereby the pipeline that produced it, on 713 endpoints from eight real-world security APIs (Table~\ref{tab:dataset}), chosen for heterogeneity in endpoint count, HTTP method distribution, authentication pattern, and request-body complexity, ranging from a two-endpoint minimal baseline (Indicators of Compromise) to the largest, most request-body-heavy API in the corpus (Wazuh). All generation experiments use the CACAO Bridge configuration from Section~\ref{sec:design}: Anthropic Claude API calls with \texttt{claude-haiku-4-5} for simple endpoints, \texttt{claude-sonnet-4-6} for endpoints with at least five parameters or a request body, default sampling temperature, and up to two schema-correction rounds.

\begin{table}[t]
\caption{Evaluation dataset: eight real-world security APIs seeding CCR.}
\label{tab:dataset}
\centering
\scriptsize
\setlength{\tabcolsep}{3pt}
\begin{tabular}{lrlr}
\toprule
\textbf{API} & \textbf{Ver.} & \textbf{Domain} & \textbf{Endpoints} \\
\midrule
CVE Services API  & 2.5.2  & Vulnerability intelligence & 25 \\
DFIR-IRIS         & v2.0   & Incident case management   & 150 \\
TheHive           & v5.7.2   & SOAR / case management     & 287 \\
Host Firewall     & v3     & Network policy control     & 18  \\
Indicators of Compromise & v3 & Threat intelligence    & 2   \\
Shodan REST API   & 1.0.0  & Internet scan / OSINT      & 45  \\
TAXII Client API  & 0.5    & CTI sharing (STIX/TAXII)   & 11  \\
Wazuh REST API    & 4.10.1   & SIEM / EDR                 & 175 \\
\midrule
\textbf{Total}    &        &                            & \textbf{713} \\
\bottomrule
\end{tabular}
\end{table}

\label{sec:backval}Schema validity is necessary but not sufficient: an envelope can be schema-conformant yet still carry the wrong HTTP method or omit required parameters. We therefore validate every envelope against its source OpenAPI operation on nine per-endpoint checks: (1)~HTTP method; (2)~request path, after normalizing OpenAPI \texttt{\{param\}} and CACAO \texttt{\_\_param\_\_:value} syntax to equivalent placeholders; (3)~parameter recall against \texttt{parameters[]} and \texttt{requestBody} fields; (4)~parameter precision; (5)~header-parameter placement; (6)~CACAO type accuracy; (7)~required/\texttt{external}-flag accuracy; (8)~authentication presence if a security requirement is declared; and (9)~base-URL match against \texttt{servers[0].url}. The overall score is the mean of these nine checks and also CCR's acceptance threshold (Section~\ref{sec:registry}). The following subsections report the results and analyse them in relation to the research questions, together with additional findings from the evaluation.

\subsection{Results}\label{sec:results}

\begin{table}[t]
\caption{Metric taxonomy: what each evidence layer actually establishes.}
\label{tab:taxonomy}
\centering
\scriptsize
\setlength{\tabcolsep}{3pt}
\begin{tabular}{>{\raggedright\arraybackslash}p{2cm}p{4.1cm}p{5.4cm}}
\toprule
\textbf{Layer} & \textbf{What it establishes} & \textbf{Evidence in this paper} \\
\midrule
1. Deterministic preservation & Facts extractable from OpenAPI with no ambiguity, including the method and declared parameters, are preserved in the generated envelope. & Checks 1, 3, 7, and 8 above; 100.0\% method accuracy, 78.6\% parameter recall, 99.1\% required-flag accuracy, and 99.9\% authentication presence (Table~\ref{tab:results}). \\
2. LLM semantic mapping & Judgment calls not fully determined by the source schema (variable naming, type enrichment, activity annotation) are reasonable. & Checks 2, 4, 5, 6 above; the RQ1 baseline comparison isolates what only the LLM stage contributes. \\
3. Runtime request conformance & The artefact, once variables are substituted, is a well-formed, dispatchable HTTP request. & Local dispatch harness, 675/713 (Section~\ref{sec:execval}); does \emph{not} establish the target system accepts it. \\
4. Live orchestration compatibility & A real target accepts the request, or a CACAO workflow consumes the artefact. & Live execution, 5/5, for one public read-only CVE operation family; a minimal CACAO playbook executes via a local harness with a mock state-changing target; no third-party orchestrator integration performed. \\
5. Human-reviewed adequacy & An independent domain expert judges the artefact operationally usable. & No expert assessment was completed in this study; consequently, no claim rests on this evidence layer. \\
\bottomrule
\end{tabular}
\end{table}

A single averaged score cannot capture all dimensions of correctness. Table~\ref{tab:taxonomy} distinguishes five evidence layers and identifies which are addressed by the nine back-validation checks. Table~\ref{tab:results} reports the headline metrics, and Table~\ref{tab:per_api} provides the per-API breakdown. The 91.5\% headline figure summarises evidence from layers 1--2. Runtime request conformance and limited live-execution evidence are reported separately for layers 3--4, while layer 5 was not assessed.
All 713 endpoints were processed with zero transient LLM API-call failures. The 32 schema-correction calls (31 for DFIR-IRIS and one for TheHive) were initiated by Stage-4 validation failures and are reported separately from transient API failures.

\begin{table}[t]
    \begin{minipage}[t]{0.44\textwidth}
    \centering
    \caption{Headline results: \\(713 endpoints and 8 APIs).}
    \label{tab:results}
    \scriptsize
    \begin{tabular}{p{4.2cm}r}
    \toprule
    \textbf{Metric} & \textbf{Value} \\
    \midrule
    CACAO schema validity    & 100.0\% \\ 
    HTTP method correctness  & 100.0\% \\ 
    Authentication presence  & 99.9\% \\ 
    Required-flag accuracy   & 99.1\% \\
    Overall average accuracy & 91.5\% \\
    Transient LLM API-call failures & 0 \\
    Schema-correction rounds & 32 \\
    Total generation cost    & \$18.60 \\
    \bottomrule
    \end{tabular}
    \end{minipage}%
\hfill
\begin{minipage}[t]{0.54\textwidth}
\centering
\caption{Per-API results.}
\label{tab:per_api}
\scriptsize
\setlength{\tabcolsep}{3pt}
\begin{tabular}{lrrrr}
\toprule
\textbf{API} & \textbf{EPs} & \textbf{Schema} & \textbf{Avg Acc.} & \textbf{Cost} \\
\midrule
CVE Services  & 25  & 100\% & 84.2\% & \$1.26 \\
DFIR-IRIS     & 150 & 100\% & 93.9\% & \$2.32 \\
TheHive       & 287 & 100\% & 86.0\% & \$2.89 \\
Host Firewall & 18  & 100\% & 95.9\% & \$0.84 \\
IoC           & 2   & 100\% & 100.0\%& \$0.07 \\
Shodan        & 45  & 100\% & 99.4\% & \$1.99 \\
TAXII         & 11  & 100\% & 93.9\% & \$0.19 \\
Wazuh         & 175 & 100\% & 96.6\% & \$9.04 \\
\midrule
\textbf{Total}& 713 & 100\% & 91.5\% & \$18.60 \\
\bottomrule
\end{tabular}
\end{minipage}
\end{table}

All 713 envelopes passed CACAO v2 schema validation. First-attempt validity was 95.5\% (681/713), and the remaining 32 envelopes passed after one schema-correction round. The corrected endpoints achieved a mean back-validation score of 90.7\%, compared with 91.5\% for the endpoints that passed on the first attempt. HTTP-method accuracy was 100.0\% across all five observed verbs, required-flag accuracy was 99.1\%, and authentication-presence accuracy was 99.9\% across six observed security patterns. The authentication-presence check records whether an authentication object exists when the source declares a security requirement; it does not evaluate header placement, token format, or credential values. The overall mean back-validation score was 91.5\%, with the lowest API-level results observed for TheHive and CVE Services, primarily because of parameter-recall penalties on large request bodies.

Per-API path-check rates are misleading if read naively: our regular expression for path-normalisation matches OpenAPI's \texttt{\{param\}} against CACAO's \texttt{\_\_param\_\_:value} only for lowercase variable names. All 93 of the corpus's path-check failures (all but one in TheHive) arise from the same root cause as the naming-convention defect in Section~\ref{sec:execval}: the pipeline preserves mixed-case parameter names (e.g., \texttt{alertId}), whereas the path-normalisation regular expression accepts only lowercase names. Consequently, equivalent representations of the same endpoint receive different path scores. The raw path-check rate (86.7\%, Table~\ref{tab:baseline}) therefore understates true path-level correctness.

The LLM stage also annotates each command with a CACAO \texttt{playbook\_activity} label \cite{cacao2023}. This was achieved for 52.7\% of endpoints: many real-world operations do not map cleanly onto an existing activity type. CCR maintains a record of whether an annotation was assigned or absent. Extending the vocabulary is ongoing work beyond the scope of this paper.

\subsection{Analysis}\label{sec:execval}

\begin{table}[t]
\caption{CACAO Bridge ($n=713$) and Rule-Based Baseline (RBB; $n=712$) back-validation results. Deltas are calculated from unrounded values.}
\label{tab:baseline}
\centering
\scriptsize
\setlength{\tabcolsep}{3pt}
\begin{tabular}{lrrr}
\toprule
\textbf{Check} & \textbf{Bridge} & \textbf{RBB} & \textbf{$\Delta$ (pp)} \\
\midrule
Schema validity                    & 100.0\%           & 100.0\%           & 0.0 \\
HTTP method                        & 100.0\%           & 100.0\%           & 0.0 \\
Path                               & 86.7\%            & 100.0\%           & $-13.3$ \\
Parameter recall                   & 78.6\%            & 99.8\%            & $-21.2$ \\
Parameter precision                & 61.8\%            & 97.1\%            & $-35.3$ \\
Header placement                   & 97.9\%            & 100.0\%           & $-2.1$ \\
Type / required-flag accuracy      & 99.9\% / 99.1\%  & 100.0\% / 100.0\% & $-0.1$ / $-0.9$ \\
Auth presence / base URL           & 99.9\% / 100.0\% & 100.0\% / 100.0\% & $-0.1$ / 0.0 \\
\midrule
\textbf{Mean back-validation score} & \textbf{91.5\%} & \textbf{99.6\%} & \textbf{$-8.2$} \\
\textbf{Activity annotation}        & \textbf{52.7\%} & \textbf{0.0\%}  & \textbf{+52.7} \\
\midrule
Cost                                & \$18.60          & \$0.00           & -- \\
\bottomrule
\end{tabular}
\end{table}

To address RQ1, we implemented a \emph{Rule-Based Baseline} (RBB) by replacing Stage~2 with the fixed-rule mapping described in Section~\ref{sec:design}. Table~\ref{tab:baseline} compares the available Bridge ($n=713$) and RBB ($n=712$) outputs. RBB equals or outperforms Bridge on each mechanical check because it preserves the corresponding OpenAPI fields directly. Some differences produced by Bridge reflect semantic transformations, such as exposing \texttt{requestBody} fields as variables or refining \texttt{string} types to \texttt{uuid}; source-agreement metrics record these transformations as deviations even when they improve the resulting description. The principal contribution of the LLM is activity annotation: Bridge assigns a CACAO \texttt{playbook\_activity} label to 52.7\% of the endpoints, while RBB assigns none. These results support a division of labour in which deterministic mapping preserves mechanical API structure and the LLM supplies bounded semantic enrichment, most clearly through controlled-vocabulary activity annotation.

%\paragraph{RQ2: Execution-Oriented Validity of the Automated Gate}\label{sec:rq2}
With respect to RQ2, back-validation and the RBB comparison assess an artefact against the \emph{same} OpenAPI document used as pipeline input. A high score therefore partly reflects source reproduction and cannot independently establish operational usability. The execution evidence supports a qualified answer. The automated gate provides an effective first layer for detecting structural and source-consistency defects, but independent execution-oriented testing remains necessary to establish dispatchability. Dispatch testing exposed two defect classes among envelopes with otherwise acceptable scores---mixed-case tokens and undeclared variable references---and 5.3\% of the corpus failed dispatch outright. The base-URL audit further showed that agreement with a source specification does not ensure that the resulting address is directly usable. Gate scores must therefore be interpreted together with dispatch and target-facing evidence, as reflected by the separate maturity levels in Section~\ref{sec:governance} and Figure~\ref{fig:maturity}. As no independent expert assessment was conducted, the present evaluation does not establish human-judged usability or determine whether expert judgements align with the 0.80 acceptance threshold.

%\paragraph{Execution Validation}\label{sec:execval}
Back-validation and the RBB comparison both compare generated artefacts against the source OpenAPI \emph{specification}, never against a real, executing system; we therefore additionally validate execution readiness on two tiers: full-corpus local dispatch validation of request construction, and limited target-facing execution through live CVE requests plus a mock state-changing playbook trace. Both tiers are fully reproducible and released alongside CCR.\footnote{\texttt{execution-validation/}}

%\textbf{Live execution.} 
The only API in the corpus with genuinely public, unauthenticated, side-effect-free (read-only \texttt{GET}) endpoints is the MITRE CVE Services API. We took the pipeline-generated command for \texttt{GET /cve/\{id\}} verbatim, substituted five real published CVE identifiers, and dispatched the resulting requests against the live production API at \texttt{cveawg.mitre.org}. All five (5/5, 100\%) returned HTTP 200 with the correct published CVE record.
%\textbf{Local dispatch validation.} 
The remaining seven APIs require vendor credentials or a live deployment, neither of which was available for this study. For these, we validated the request-\emph{construction} itself. A harness substitutes synthetic values, assembles headers/auth, and dispatches each request to a local recording server, treating undeclared variables as failures. Five deliberately invalid synthetic envelopes\footnote{Malformed command, undeclared variable, missing authentication, unsupported method, and unsafe non-HTTP(S) target.} were all rejected, demonstrating that the harness detects these predefined failure modes.
Run across {all 713} seed envelopes, {675/713 (94.7\%)} produced well-formed, dispatchable requests. The 38 failures exposed two defect classes missed by back-validation: {94/713 (13.2\%)}, all in TheHive, carried mixed-case variable tokens\footnote{7 fail dispatch; 87 substitute but violate strict CACAO naming.}; and {31/713 (4.3\%)} reference variables with no matching \texttt{step\_variables} declaration, typically bearer-token or path identifiers, named inconsistently between declaration and use.

%\textbf{A third finding: base-URL usability.} 
461/713 (64.7\%) of envelopes carry a target address not directly usable as an absolute URL, a distinction the back-validation base-URL check cannot make, since it only compares against what the OpenAPI specification declares. TheHive (286/287) inherits the literal string \texttt{/} from its OpenAPI document, reflecting a limitation that cannot be resolved from the specification alone. Wazuh (175/175) inherits the unresolved template \texttt{\{protocol\}://\{host\}:\{port\}}, although its specification declares usable defaults (\texttt{https}, \texttt{localhost}, and \texttt{55000}). These defaults could be used by a future pipeline stage to resolve the template.

\label{sec:variance}
%\paragraph{Output Consistency Across Repeated Runs}\label{sec:variance}
No explicit sampling temperature was specified for the LLM calls; consequently, the provider's default configuration was used and variation between repeated generations was possible. To assess this variation, we selected 22 endpoints stratified by API, with two or three endpoints drawn from each API, and generated each endpoint five times, yielding 110 additional calls.\footnote{\texttt{execution-validation/variance\_study.py}} All 110 outputs passed CACAO schema validation, indicating stable structural validity within the evaluated sample. Back-validation scores varied across repetitions for 12 of the 22 endpoints (54.5\%). The mean endpoint-level standard deviation was 0.025, with a median of 0.015 and a maximum of 0.076. These results indicate that semantic back-validation scores exhibit run-to-run variation even when schema validity remains unchanged. The endpoint-level and corpus-level scores reported from the main experiment should therefore be interpreted as single-run estimates. A larger repeated-run evaluation would be required to quantify their uncertainty more precisely.

\section{Discussion and Related Work}\label{sec:discussion}
The present work sits at the intersection of cybersecurity playbook standardisation, machine-readable API specifications, and open cybersecurity infrastructure policy. Prior CACAO work covers playbook sharing in threat-intelligence exchange, semantic-web-based management, and lifecycle knowledge management \cite{mavroeidis2021integration,gurabi2022sasp,tsirakis2025operationalizing}; it establishes CACAO as operationally important but does not automate the vendor-specific integrations a playbook needs to run.

The closest prior work is APIRO \cite{sworna2023apiro} and IRP2API \cite{sworna2023irp2api}. APIRO recommends the top-$k$ most relevant APIs for a task from a fixed candidate pool, while IRP2API maps incident-response-plan tasks to security-tool APIs. Both address API retrieval and report ranking metrics (Top-$k$, MRR), but neither constructs the authenticated request, parameter bindings, and target configuration required by an orchestrator. This paper addresses the subsequent integration stage by generating and validating an executable CACAO binding for a selected API operation. Combining this capability with IRP2API-style recommendation would support a broader workflow from API selection to executable integration.

Structured, non-LLM playbook-generation approaches include ICS vulnerability playbooks from CSAF advisories \cite{empl2024ics}, model-based playbook design \cite{shaked2022model}, and attack-graph/cloud-native remediation \cite{sainthilaire2025graphs,leone2025remediation}, all with substantial knowledge-engineering overhead. Recent LLM-based CACAO work targets \emph{text-to-playbook} transformation rather than integration synthesis \cite{paduraru2025automated,gurabi2025legacy}. These approaches address adjacent aspects of playbook generation but do not cover the executable-integration layer considered here.

REST/OpenAPI software-engineering research treats specifications as engineering assets \cite{casas2021openapi,peldszus2026developerperspectivesrestapi,LERCHER2024112110}, supporting OpenAPI as a grounding source but not addressing CACAO semantics. Tool-augmented LLM research shows that LLMs can invoke and parameterise APIs under schema constraints \cite{yao2023react,schick2023toolformer,patil2023gorilla,li2023apibank,qin2024toolllm,scholak2021picard}, motivating constrained-generation design without addressing standards-grounded orchestration or open infrastructure policy.

Structurally, OpenAPI-to-CACAO translation contains a \emph{schema-matching} subproblem: establishing correspondences between independently designed data models \cite{rahm2001schemamatching}. Matching alone is not a translation, because an executable CACAO binding also requires orchestration semantics, target construction, authentication objects, variable externalization, and schema-valid cross-references. Schema matching and \emph{ontology alignment} nevertheless formalize the evaluation problem for the correspondence layer: scoring precision/recall-style agreement rather than exact equality, and distinguishing beneficial enrichment from structural error \cite{euzenat2013ontologymatching}. Our back-validation protocol applies this paradigm to OpenAPI-to-CACAO mapping and provides a methodological basis for evaluating correspondence between the two representations. To our knowledge, no prior work combines (i)~AI-assisted OpenAPI-to-CACAO generation evaluated with schema-matching-style metrics, (ii)~an openly licensed, quality-gated registry of the results, and (iii)~explicit grounding in EU cybersecurity regulation as the deployment context.

%\subsection{Implications}
The observed model-inference cost of \$0.026 per endpoint is low, but excludes subsequent review and governance costs. The 91.5\% mean back-validation score should also be interpreted in light of the case-sensitive path-normalisation defect, which affects both the evaluator and generated artefacts and requires separate corrections to each. Applying the same gate to AI-generated and manually authored contributions makes acceptance independent of the production method, supporting a multi-contributor EU-wide registry.

%\subsection{Limitations and Threats to Validity}
We structure the following limitations according to the standard empirical software-engineering validity taxonomy \cite{wohlin2012experimentation}.
\textit{Construct validity.} The back-validation path-normalisation regular expression is case-sensitive and silently fails on mixed-case variable names. The same defect is quantified in Section~\ref{sec:variance}; these results therefore describe one defect class, not two independent findings. More fundamentally, back-validation and the RBB comparison both assess an artefact against the \emph{same} OpenAPI document used as pipeline input. Execution validation partially mitigates this threat by testing request construction independently of source-field agreement. However, independent expert assessment was not completed, limiting the conclusions that can be drawn about human-judged operational adequacy.

\textit{Internal validity.} The headline results in Section~\ref{sec:results} are derived from a single generation run. The pipeline uses the model's default sampling temperature, so stochastic variation is expected; Section~\ref{sec:variance} characterises this variation on a repeated-run sample. The schema-correction loop affects the final schema-validity result. We therefore report first-pass validity separately: 681/713 envelopes (95.5\%) passed before correction, increasing to 713/713 (100\%) after correction. This distinction prevents the final validity rate from being interpreted as the result of a single generation attempt. The artefact package supports inspection and reproduction of the registry artefacts, validation results, and execution traces. Full regeneration of the corpus is not possible because the prototype generator has not been released.

\textit{External validity.} The seed corpus contains 713 endpoints from eight security APIs, selected to vary in endpoint count, HTTP-method distribution, authentication pattern, request-body complexity, and operational domain. This variation supports evaluation across several types of OpenAPI operation, but does not establish representativeness of security APIs more generally. The scope is limited to HTTP APIs described with OpenAPI 3.x; GraphQL, gRPC, and proprietary protocols used in industrial and operational technology environments are not represented. The APIs form a convenience sample and do not constitute a stratified sample of the wider API population. Observed proportions, including the 13.2\% naming-convention defect rate, demonstrate the presence of a defect class but should not be interpreted as population-level estimates. Although previous work describes manual connector authoring as resource-intensive \cite{sworna2023apiro,islam2020architecture}, no quantified person-hours baseline applicable to this corpus is available. Section~\ref{sec:results} therefore reports generation cost without estimating an effort-reduction ratio. A timed study with professional integration developers is required to quantify that effect.

\textit{Conclusion validity.} CCR implements assurance metadata, but its governance process has not been evaluated at scale. Adoption by external contributors and the robustness of the quality gate against adversarial submissions require evaluation in a publicly operating registry. The mapping to NIS2, CER, and the CRA is limited to provisions with a direct technical relationship to the registry and does not establish legal compliance.

\section{Conclusion}\label{sec:conclusion}
This paper introduced the Common CACAO Registry (CCR), an open repository of quality-gated CACAO playbook integration candidates represented under a common standard and initially seeded with candidates generated from vendor OpenAPI specifications. CCR addresses a technical integration barrier relevant to the incident-response and resilience capabilities promoted by NIS2, CER, and the CRA. The regulatory discussion is based on a scoped mapping between requirements and technical capabilities, while the open licensing and contribution model support European objectives for interoperability, reuse, and digital autonomy.

The evaluation clarifies the appropriate division of labour in OpenAPI-to-CACAO translation. The rule-based mapper equals or exceeds the LLM pipeline on all mechanical checks, supporting deterministic preservation of source-derived interface facts. The principal contribution of the LLM stage is bounded semantic enrichment, especially vocabulary-grounded activity annotation (RQ1). The automated quality gate identifies structural and source-consistency defects, while full-corpus dispatch testing revealed two additional defect classes and a 5.3\% dispatch-failure rate. Live execution, the minimal playbook demonstration, and the output-consistency study further delimit the claims supported by a single back-validation score. CCR therefore records gate results and execution evidence at separate maturity levels (RQ2). The six-level maturity model and assurance-envelope schema provide a common structure for accumulating stronger evidence as the registry develops. An independent expert assessment was not completed in this study; consequently, no reported result or maturity claim relies on expert review.

Future work will conduct an expert-rating study and add normalised path comparison to the back-validator. Further evaluation will cover the remaining seven APIs when sandboxed deployments become available, execute the demonstration playbook through SOARCA, and conduct a governance trial with external contributors. A companion study is also investigating extensions to CACAO's activity-type vocabulary based on the coverage gaps observed during CCR seeding.

\begin{credits}
\subsubsection{Data and Artefact Availability}
The CCR artefact package contains the 713 seed envelopes, source OpenAPI files with SHA-256 hashes, assurance metadata, validation harnesses, raw results, the demonstration playbook, its captured trace, and the expert-rating materials. The prototype generator is not included. Expert-rating results will be added upon completion of the study.

\subsubsection{\ackname}
The authors would like to thank the OASIS CACAO Technical Committee and the broader open-standards community for their contributions to interoperable cybersecurity playbooks.

%\subsubsection{\discintname}
%The authors have no competing interests to declare that are relevant to the content of this article.
\end{credits}

\bibliographystyle{splncs04}
\bibliography{references}

@article{islam2019multivocal,
  author = {Islam, Chadni and Babar, Muhammad Ali and Nepal, Surya},
  title = {A Multi-Vocal Review of Security Orchestration},
  journal = {ACM Comput. Surv.},
  volume = {52},
  number = {2},
  pages = {1--45},
  year = {2019},
  doi = {10.1145/3305268}
}

@techreport{cacao2023,
  author = {Jordan, Bret and Thomson, Allan},
  title = {{CACAO} Security Playbooks Version 2.0},
  institution = {{OASIS}},
  type = {Committee Spec. 01},
  year = {2023},
  month = nov,
  url = {https://docs.oasis-open.org/cacao/security-playbooks/v2.0/cs01/security-playbooks-v2.0-cs01.html}
}

@misc{openapi2021,
  author = {{OpenAPI Initiative}},
  title = {{OpenAPI} Specification v3.1.0},
  howpublished = {{OpenAPI} Initiative},
  year = {2021},
  month = feb,
  url = {https://spec.openapis.org/oas/v3.1.0.html}
}

@article{sworna2023apiro,
  author = {Sworna, Zarrin Tasnim and Islam, Chadni and Babar, Muhammad Ali},
  title = {{APIRO}: A Framework for Automated Security Tools {API} Recommendation},
  journal = {ACM Trans. Softw. Eng. Methodol.},
  volume = {32},
  number = {1},
  pages = {1--42},
  year = {2023},
  doi = {10.1145/3512768}
}

@inproceedings{sworna2023irp2api,
  author = {Sworna, Zarrin Tasnim and Babar, Muhammad Ali and Sreekumar, Anjitha},
  title = {{IRP2API}: Automated Mapping of Cyber Security Incident Response Plan to Security Tools' {APIs}},
  booktitle = {Proc. {IEEE SANER}},
  pages = {546--557},
  year = {2023},
  doi = {10.1109/SANER56733.2023.00057}
}

@inproceedings{islam2020architecture,
  author = {Islam, Chadni and Babar, Muhammad Ali and Nepal, Surya},
  title = {Architecture-Centric Support for Integrating Security Tools in a Security Orchestration Platform},
  booktitle = {Proc. {ECSA}},
  series = {{LNCS}},
  volume = {12292},
  pages = {165--181},
  publisher = {Springer},
  year = {2020},
  doi = {10.1007/978-3-030-58923-3_11}
}

@inproceedings{mavroeidis2021integration,
  author = {Mavroeidis, Vasileios and Eis, Pavel and Zadnik, Martin and Caselli, Marco and Jordan, Bret},
  title = {On the Integration of Course of Action Playbooks into Shareable Cyber Threat Intelligence},
  booktitle = {Proc. {IEEE Big Data}},
  pages = {2104--2108},
  year = {2021},
  doi = {10.1109/BigData52589.2021.9671893}
}

@inproceedings{gurabi2022sasp,
  author = {Akbari Gurabi, Mehdi and Mandal, Avikarsha and Popanda, Jan and Rapp, Robert and Decker, Stefan Josef},
  title = {{SASP}: A Semantic Web-Based Approach for Management of Sharable Cybersecurity Playbooks},
  booktitle = {Proc. {ARES}},
  pages = {1--8},
  year = {2022},
  doi = {10.1145/3538969.3544478}
}

@article{tsirakis2025operationalizing,
  author = {Tsirakis, Orestis and Fysarakis, Konstantinos and Mavroeidis, Vasileios and Papaefstathiou, Ioannis},
  title = {Operationalizing Cybersecurity Knowledge: Design, Implementation \& Evaluation of a Knowledge Management System for {CACAO} Playbooks},
  journal = {Comput. Secur.},
  volume = {159},
  pages = {104696},
  year = {2025},
  doi = {10.1016/j.cose.2025.104696}
}

@article{empl2024ics,
  author = {Empl, Philip and Schlette, Daniel and St\"{o}ger, Lukas and Pernul, G\"{u}nther},
  title = {Generating {ICS} Vulnerability Playbooks with Open Standards},
  journal = {Int. J. Inf. Secur.},
  volume = {23},
  pages = {1215--1230},
  year = {2024},
  doi = {10.1007/s10207-023-00760-5}
}

@inproceedings{shaked2022model,
  author = {Shaked, Avi and Cherdantseva, Yulia and Burnap, Pete},
  title = {Model-Based Incident Response Playbooks},
  booktitle = {Proc. {ARES}},
  pages = {1--7},
  year = {2022},
  doi = {10.1145/3538969.3538976}
}

@phdthesis{sainthilaire2025graphs,
  author = {Saint-Hilaire, Keren A.},
  title = {Automatic Generation of Attack and Remediation Graphs},
  school = {Polytechnique Montr{\'e}al},
  year = {2025},
  url = {https://publications.polymtl.ca/63355/}
}

@mastersthesis{leone2025remediation,
  author = {Leone, Dario Simone},
  title = {Remediation Procedures and Automated Cybersecurity Incident Response},
  school = {Politecnico di Torino},
  year = {2025},
  url = {https://webthesis.biblio.polito.it/id/eprint/37719/}
}

@article{paduraru2025automated,
  author = {Paduraru, Ciprian and Dumitru, Bogdan and Stefanescu, Alin},
  title = {Automated Generation of Cybersecurity Response Playbooks via Large Language Models},
  journal = {Procedia Comput. Sci.},
  volume = {270},
  pages = {2987--2996},
  year = {2025},
  doi = {10.1016/j.procs.2025.09.423}
}

@misc{gurabi2025legacy,
  author = {Akbari Gurabi, Mehdi and Nitz, Lasse and Castravet, Radu-Mihai and Matzutt, Roman and Mandal, Avikarsha and Decker, Stefan},
  title = {From Legacy to Standard: {LLM}-Assisted Transformation of Cybersecurity Playbooks into {CACAO} Format},
  year = {2025},
  doi = {10.48550/arXiv.2508.03342},
  archivePrefix = {arXiv},
  eprint = {2508.03342},
  primaryClass = {cs.CR}
}

@misc{patil2023gorilla,
  author = {Patil, Shishir G. and Zhang, Tianjun and Wang, Xin and Gonzalez, Joseph E.},
  title = {Gorilla: Large Language Model Connected with Massive {APIs}},
  year = {2023},
  doi = {10.48550/arXiv.2305.15334},
  archivePrefix = {arXiv},
  eprint = {2305.15334},
  primaryClass = {cs.LG}
}

@misc{li2023apibank,
  author = {Li, Minghao and Song, Feifan and Yu, Bowen and Yu, Haiyang and Li, Zhoujun and Huang, Fei and Li, Yongbin},
  title = {{API}-Bank: A Benchmark for Tool-Augmented {LLMs}},
  year = {2023},
  doi = {10.48550/arXiv.2304.08244},
  archivePrefix = {arXiv},
  eprint = {2304.08244},
  primaryClass = {cs.CL}
}

@inproceedings{casas2021openapi,
  author = {Casas, Sandra and Cruz, Diana and Vidal, Graciela and Constanzo, Marcela Alejandra},
  title = {Uses and Applications of the {OpenAPI/Swagger} Specification: A Systematic Mapping of the Literature},
  booktitle = {Proc. {IEEE SCCC}},
  pages = {1--8},
  year = {2021},
  doi = {10.1109/SCCC54552.2021.9650408}
}

@inproceedings{yao2023react,
  author = {Yao, Shunyu and Zhao, Jeffrey and Yu, Dian and Du, Nan and Shafran, Izhak and Narasimhan, Karthik and Cao, Yuan},
  title = {{ReAct}: Synergizing Reasoning and Acting in Language Models},
  booktitle = {{ICLR}},
  year = {2023},
  url = {https://openreview.net/forum?id=WE_vluYUL-X}
}

@inproceedings{schick2023toolformer,
  author = {Schick, Timo and Dwivedi-Yu, Jane and Dessi, Roberto and Raileanu, Roberta and Lomeli, Maria and Zettlemoyer, Luke and Cancedda, Nicola and Scialom, Thomas},
  title = {Toolformer: Language Models Can Teach Themselves to Use Tools},
  booktitle = {{NeurIPS}},
  volume = {36},
  pages = {68539--68551},
  year = {2023},
  url = {https://proceedings.neurips.cc/paper_files/paper/2023/hash/d842425e4bf79ba039352da0f658a906-Abstract-Conference.html}
}

@inproceedings{qin2024toolllm,
  author = {Qin, Yujia and Liang, Shihao and Ye, Yining and Zhu, Kunlun and Yan, Lan and Lu, Yaxi and Lin, Yankai and Cong, Xin and Tang, Xiangru and Qian, Bill and Zhao, Sihan and Hong, Yansong and Tian, Ruobing and Xie, Ruobing and Zhou, Jie and Gerstein, Mark and Li, Dahai and Liu, Zhiyuan and Sun, Maosong},
  title = {{ToolLLM}: Facilitating Large Language Models to Master 16000+ Real-World {APIs}},
  booktitle = {{ICLR}},
  year = {2024},
  url = {https://openreview.net/forum?id=dHng2O0Jjr}
}

@inproceedings{scholak2021picard,
  author = {Scholak, Torsten and Schucher, Nathan and Bahdanau, Dzmitry},
  title = {{PICARD}: Parsing Incrementally for Constrained Auto-Regressive Decoding from Language Models},
  booktitle = {Proc. {EMNLP}},
  pages = {9895--9901},
  year = {2021},
  doi = {10.18653/v1/2021.emnlp-main.779}
}

@INPROCEEDINGS {9240687,
author = { Yasmin, Jerin and Tian, Yuan and Yang, Jinqiu },
booktitle = {Proc. {IEEE ICSME}},
title = {{ A First Look at the Deprecation of RESTful APIs: An Empirical Study }},
year = {2020},
volume = {},
ISSN = {},
pages = {151-161},
doi = {10.1109/ICSME46990.2020.00024},
month =Oct}

@article{LERCHER2024112110,
title = {Microservice {API} Evolution in Practice: A Study on Strategies and Challenges},
journal = {J. Syst. Softw.},
volume = {215},
pages = {112110},
year = {2024},
doi = {10.1016/j.jss.2024.112110},
author = {Alexander Lercher and Johann Glock and Christian Macho and Martin Pinzger}
}

@misc{peldszus2026developerperspectivesrestapi,
      title={Developer Perspectives on {REST API} Usability: A Study of {REST API} Guidelines},
      author={Sven Peldszus and Jan Rutenkolk and Marcel Heide and Jan Sollmann and Benjamin Klatt and Frank Köhne and Thorsten Berger},
      year={2026},
      doi={10.48550/arXiv.2601.16705}
}

@misc{nis2directive,
  author = {{European Parliament and Council}},
  title = {{NIS 2 Directive}: Directive ({EU}) 2022/2555},
  howpublished = {Off. J. Eur. Union, L 333},
  year = {2022},
  url = {https://eur-lex.europa.eu/eli/dir/2022/2555/oj/eng}
}

@misc{cerdirective,
  author = {{European Parliament and Council}},
  title = {{CER Directive}: Directive ({EU}) 2022/2557},
  howpublished = {Off. J. Eur. Union},
  year = {2022},
  url = {https://eur-lex.europa.eu/eli/dir/2022/2557/oj/eng}
}

@misc{craregulation,
  author = {{European Parliament and Council}},
  title = {{Cyber Resilience Act}: Regulation ({EU}) 2024/2847},
  howpublished = {Off. J. Eur. Union},
  year = {2024},
  url = {https://eur-lex.europa.eu/eli/reg/2024/2847/oj/eng}
}

@article{rahm2001schemamatching,
  author = {Rahm, Erhard and Bernstein, Philip A.},
  title = {A Survey of Approaches to Automatic Schema Matching},
  journal = {VLDB J.},
  volume = {10},
  number = {4},
  pages = {334--350},
  year = {2001},
  doi = {10.1007/s007780100057}
}

@book{wohlin2012experimentation,
  author = {Wohlin, Claes and Runeson, Per and H\"{o}st, Martin and Ohlsson, Magnus C. and Regnell, Bj\"{o}rn and Wessl\'{e}n, Anders},
  title = {Experimentation in Software Engineering},
  publisher = {Springer},
  address = {Berlin Heidelberg},
  year = {2012},
  doi = {10.1007/978-3-642-29044-2}
}

@book{euzenat2013ontologymatching,
  author = {Euzenat, J\'{e}r\^{o}me and Shvaiko, Pavel},
  title = {Ontology Matching},
  edition = {2nd},
  publisher = {Springer},
  address = {Heidelberg},
  year = {2013},
  doi = {10.1007/978-3-642-38721-0}
}

@misc{cybersolidarityact,
  author = {{European Parliament and Council}},
  title = {{Cyber Solidarity Act}: Regulation ({EU}) 2025/38},
  howpublished = {Off. J. Eur. Union},
  year = {2025},
  url = {https://eur-lex.europa.eu/eli/reg/2025/38/oj/eng}
}

@misc{interoperableeuropeact,
  author = {{European Parliament and Council}},
  title = {{Interoperable Europe Act}: Regulation ({EU}) 2024/903},
  howpublished = {Off. J. Eur. Union},
  year = {2024},
  url = {https://eur-lex.europa.eu/eli/reg/2024/903/oj/eng}
}

@misc{enisaeuvd,
  author = {{ENISA}},
  title = {European Vulnerability Database ({EUVD})},
  howpublished = {{ENISA}},
  year = {2025},
  url = {https://euvd.enisa.europa.eu/}
}

@misc{horizonccl3eccc2026,
  author = {{ECCC}},
  title = {{HORIZON-CL3-2026-02-CS-ECCC-01}: Security Development and Assessment},
  howpublished = {Horizon Europe WP 2026--2027},
  year = {2026},
  url = {https://cordis.europa.eu/programme/id/HORIZON_HORIZON-CL3-2026-02-CS-ECCC-01}
}

@misc{horizoncl4openinternetstack2026,
  author = {{EC}},
  title = {{HORIZON-CL4-2026-04-DATA-02}: Open Internet Stack Sovereign Solutions},
  howpublished = {Horizon Europe WP 2026--2027},
  year = {2026},
  url = {https://topictree.ideal-ist.eu/topic/7109}
}

@misc{enisa2025threatlandscape,
  author       = {{ENISA}},
  title        = {{ENISA Threat Landscape 2025}},
  year         = {2025},
  organization = {European Union Agency for Cybersecurity},
  url          = {https://www.enisa.europa.eu/publications/enisa-threat-landscape-2025},
  note         = {Version 1.2, revised 9 January 2026. Accessed 15 August 2026}
}

\end{document}